\documentclass[aps,pre,floatfix,nofootinbib,showpacs,amsmath,amssymb]{revtex4}

\usepackage{times}
\usepackage[sort&compress]{natbib}
\usepackage[dvips]{graphicx}
\usepackage[dvips]{color}
\begin{document}

\title{  The cell cycle related rhythms, cells' states correlation  and the cancer}

\author{David B. Saakian$^{1,2}$}
\email{saakian@yerphi.am}
 \affiliation{$^2$Institute of Physics,
Academia Sinica, Nankang, Taipei 11529, Taiwan} \affiliation{$^1$
A.I. Alikhanyan National Science Laboratory (Yerevan Physics
Institute) Foundation,\\
 2 Alikhanian Brothers St., Yerevan 375036, Armenia}

\begin{abstract} We consider the decision making by mammalian cells,
looking them as dynamic systems with rhythms. We calculate the
effective dimension of the cell division model of the healthy
mammalian cells consistent with the data: it is described via a
four dimensional dynamic system.
 We assume that the  cell's decision making property is strongly
affected by the cells rhythms, their causal relations, and by the
correlation between internal states of different cells in tissue.
There is a strong correlation between the states of different
healthy cells (verified partially experimentally), and we assume
that there is no such a correlation between internal states of the
healthy and cancer cells. The origins of the cancer are just the
disruption of this correlation (self-identification of the cells)
and the change of the causal relations between the Circadian and
cell cycle rhythms. Assuming the Gaussian channel version of the
cell-cell communications and a key role of public goods for the
cancer cells, we get a strong correlation between the states of
different cancer cells.


\end{abstract}

\maketitle

\maketitle

\section{Introduction}
Among the key features of cancer the wrong division process of the
 cells  [1] and evolutionary dynamics aspects of the cell
 proliferation  [2] are well known.
 In the recent paper \cite{me15}, it has discussed cancer as a
 breaking of cooperation in multicellularity, and looked at 5 aspects: cheating
in proliferation inhibition, cell death, division of labor,
resource allocation, and extracellular environment maintenance.
Why does the normal multi-cellularity fail?
 According to
[4-7]  the (bacterial) cell has an internal representation of the
environment. It is connected somehow with the dynamics of
regulatory networks. It is possible to estimate even the
information capacity of the cells, and in the case of bacteria, it
has been estimated as a few bits [8]. In [9] we found different
phases in the dynamics of the reaction networks and related the
origin of the cancer with the change of the dynamic phases of
reaction networks.
 In recent
experimental work [10],  the statistics of the healthy mammalian
cell division times  have been analyzed by measuring the L1210
lymphoblasts division periods for different off springs of the
same cell. They fitted the data by a deterministic model and
concluded that the healthy cell's division dynamics hasve an
effective dimension of three.

In L. Schwartz et al, "Cell proliferation as the consequence of a
balance between Pentose Phosphate Pathway and Mitochondrial
Metabolism,"  several oscillations (ATP, NADH/NAD+, Ph
concentrations) have been found in the case of the healthy colon
cells, while absent in the cancer case. We interpret these
oscillations as related with the internal state of the cell and
sensing abilities of the cells allowing proper multi-cell decision
making, and explain the cancer as a lack of the mentioned
abilities. In  [11]  the common sensing of the concentration
gradients has been investigated, and it has been concluded that a
strong correlation exists between cells related to each other by
3-4 neighbor contacts. In [11] the authors considered the case of
the relay channel [12] for information processing of $n$ cells
located on a 1-d chain, and its environment. Fig.~1.a describes
the case $n=2$. Cell A senses the environment and sends a signal
to the cell B, which also gets a signal from the environment as
well. In [13],  the correlation of the calcium fluctuations in
different mammalian cells on 2-d surface have been investigated
including some fraction of cancer cells. We will correct the
mathematical result of [10], then give our interpretation of the
results [11],[13].

\section{The cell as four dimensional dynamic system during
the division process}

In (\cite{sa15}  the following system of iterative equations has
been suggested to define  $T_n$, the duration of the cell division
period after n divisions:
\begin{eqnarray}
T_{n+1}=T_0(1-\alpha)+T_n+\sin(2\pi
t_{n+1})+\epsilon^{\pm}\nonumber\\
t_{n+1}=T_n+t_n\nonumber\\
\epsilon^{\pm}=\beta (X_{n+1}^{\pm}-Y_{n+1}^{\pm})\nonumber\\
X_{n+1}^{\pm}=2(X_n\pm \delta)+(Y_n\pm \delta) Mod(1)\nonumber\\
y_{n+1}^{\pm}=(X_n\pm \delta)+(Y_n\pm \delta) Mod(1)
\end{eqnarray}
Here $T_n,t_n, X_n,Y_n$ are dynamical variables, and this is
4-dimensional dynamic system. $t_n$ is the absolute time,
$X_n,Y_n$ are introduced to describe the fluctuations of the cell
division period $T_n$ while $\alpha,\beta,\delta, T_0$ are the
parameters of the model.
 The model fits experimental data well for the correlation of division periods for
different offsprings of the same cell. It describes a map kicked
by the oscillator, plus a small term, derived using a highly
chaotic cat map for $X_n,Y_n$.

  We can remove $t_n$, then in
the argument of $Sin$ we will get a sum $T_n + ..T_1$, which is a
strongly non-Markovian behavior. In [10] the Grassberger-Procaccia
algorithm has been used to calculate the effective dimension of a
dynamic system. One puts the points ($T_n, T_{n+1},... T_{n+d})$
into d-dimensional "embedding" space, then fits the number of
these points in a d-dimensional sphere $N(r)$ as some degree of
$r$ (the distance from some reference point), such that $N(r)=c
r^{n(d)}$. Then $n(d)$ is just the correlation dimension. $n(d)$
grows with $d$, and for finite dimensional dynamic  systems goes
to a plateau at high values of d. We denoted by $D$ the limiting
value of $n(d)$. $D$ is the effective dimension of dynamical
system, characterizing its attractor dimension. The behavior of
$n(d)$, does not carry any information about the dynamical system,
only the plateau value $D$ is important. We have done numerics for
$10^8$ points, fitting the correlation dimension at the middle
scales. We took the following values for the parameters: $T_0 =
0.75, \alpha = 0.5, k = 0.1, \delta = 0.15, \beta = 0.12$,
measuring the time in days [10]. Then we get the results given by
$n_2$ in Table I while the results of [10] are denoted by $n_1$.
Our results support the value $D=4$. The correct value of
effective dimension is important for the applications. Thus,
according to experimental results of [10], the cell could be
considered as a deterministic 4 dimensional system. In [10] a
wrong result, $D=3$, has been derived as a consequence of  the
short number of iterations and of the measuring the distribution
at a small distance. We replaced the cat map by tend, Hennon or
logistic maps and obtained that the effective dimension is around
2. Therefore, these maps cannot describe the data of [10]. The
choice of the correct model, done in [10], is a really genuine
finding.

\begin{table}[tbhp]
\caption{The correlation dimensions $n_1$ from [10] and our
numerical result $n_2$ versus the embedding dimension d. The
maximal value of $n_2$ is close to 4, an effective dimension of
the system.}

\begin{tabular}{l@{\hspace{4mm}}c@{\hspace{4mm}}c@{\hspace{4mm}}c}
\hline\hline  d &2 &3 &4\\
\hline\hline  $n_1$ & 1.65 &2.05 &2.1\\
\hline\hline  $n_2$ & 1.91 &2.99 &3.93\\
 \hline\hline
\end{tabular}
\end{table}

\section{The decision making by the cells}

\subsection{ The decision making scheme} During its lifetime a
cell changes the phase of mitosis, cooperates with other cells and
reacts to the extracellular stimuli. All of these processes are
predefined by the genetic code. However, the cells do not simply
follow instructions, but they are able to make decisions based on
the received information [14].
 We can
concretize the decision making scheme by looking at the states
(internal images), sensing, and action [14]. Additionally, we
identify the cell states with the attractor points of the dynamics
of the reaction networks or gene regulation networks. The
transitions between the attractors correspond to the change of
phenotypes [15], and tradeoffs. The cells can construct their
internal representations of the environment by using the memory of
the regularity networks, similar to the case of neural networks
[4].

\begin{table}
\caption{The cell state with different level of coarse-graining
and   decision making.} \label{tab.1}
\begin{center}
\begin{tabular}{llcr}
Level & \\
\hline\hline
& Decision making\\
\hline\hline 4 & Collective rhythms, their causal\\
&relations\\
\hline\hline  3 & The environment's images, regulation \\
&networks' states, information\\
& processing  \\
\hline\hline  2 & Metabolism, polarization, pressure  \\
\hline\hline  1 & Un-equilibrium statistical mechanics. \\
 \hline\hline
\end{tabular}
\end{center}
\end{table}

In Table II we suggested a hierarchical  scheme while looking at
different aspects of decision making. There are more
deterministic, physical factors, like metabolism, polarization,
pressure and stiffness, as well as more probabilistic factors
related to information processing.  We assume as a hypothesis
({\bf Hypothesis I)} that the decision making by the cells is
defined by some key rhythms, their correlation and casual
relations. The causal relation between Circadian and cell-cycle
rhythms has been investigated in [16] for the concrete version of
mammalian cells. According to [17], the choice of pacemaker
(circadian rhythm or cell-cycle) depends on the concrete case of
the cell.

\subsection{ The rhythm as  a compressed description of the complex
system}

In living systems, we see a specific form of non-equilibrium
statistical mechanics (metabolism) and information processing
[18]. Then we add to this scheme the collective rhythms, their
correlations and causal relations. Our suggestion assumes a more
reliable scheme of decision making than the direct influence of
metabolism and information processing.
 The phases of the rhythms in living system,
and the correlations between rhythms give examples of compressed
information. Gell-Mann and Lloyd [19] defined compressed
information as a key feature of the complex adaptive system, see
also [20]. The synchronization of different rhythms is an
important characteristic of health which is well known in
biomedicine data analysis
[21,22].
 We put the decision making property at the
top of the hierarchy (see [20] for the hierarchy of "reflections"
in complex systems) levels, higher than the information
processing.

The rhythms give a simple description of the dynamics, everything
is almost the same besides the single variable  "phase" which is
not changed. In quantum mechanics, the state of the system simply
rotates in the Hilbert space while during the measurement there is
an irreversible event, a collapse of wave function. The
irreversible event of cell division is equivalent to the collapse
of the wave function. In quantum mechanics, there is also a
property like decision making due to dual property: the same
object is both a particle and a wave.
 The
causal relations between rhythms resembles a dual aspect of
quantum-mechanical systems. We need at least two "independent"
oscillations to have this quantum mechanical analogy and decision
making, like in mutation-selection pair in evolution models, to
get a quantum-mechanical mathematical structure in evolutionary
dynamics [20].

\section{The correlation between the internal states of the cells}
\subsection{The correlation}
 Let us first define the state of the cell
more accurately, see Table III. The concept of the state includes
both the steady state of gene regulation network GRN, and the
dynamic memory related to the GRN bi-stabilities
 as has been discussed for the case of bacteria
[4,5,6]. To understand the multi-cellurar phenomenon, we should
look at the sensing environment (chemosensing, mechanosensing,
photosensing), as well as the cell-cell interactions
(communications).
 The mammalian cells can sense collectively
much stronger than separately [23].  Due to fast information
transition between neighboring cells, they are undergoing a
"coherent" signal detection by several cells, suppressing the
noise-signal ratio. Actually, the fast communication amongst the
cells allows collective information processing for $n=3-4$ layers
of cells, suppressing noise-signal ratio threshold $n^2$ times
[11]. In [13] the correlation of Ca concentration's fluctuations
in the ensemble of mammalian cells (mouse fibroblast NIH 3T3
cells)has been analyzed including some fraction of the cancer
cells (MDA-MB-231).

\begin{table}
\caption{The state of the cells.} \label{tab.3}
\begin{center}
\begin{tabular}{llcr}
Level & \\
\hline\hline  3 &  The phases of the rhythms \\
\hline\hline  2 & Internal images of the environment.   \\
\hline\hline  1 & The  dynamic memory of gene \\
&regulation network \\
\hline\hline  0 & The steady state of gene regulation network \\
 \hline\hline
\end{tabular}
\end{center}
\end{table}

We assume {\bf Hypothesis II}: there is a correlation between
internal states of different healthy mammalian cells in tissue,
rhythms, and fluctuations. The correlation of internal images of
the cells allows common information processing and  therefore also
 common decision making by a collection of the cells. Without the
former, the latter certainly could not work efficiently.

 For the
diffusive molecule, it should be easier to go from one molecule to
a neighboring one when they oscillate coherently, as it is easier
to jump from one moving car to another one, moving in parallel.
The correlation between the states of the cells can be organized
both by pressure and communication between the cells. In [13], the
cell-cell communication (diffusion) rate via a gap junction has
been found to be much weaker in the cancer case, typically
30\%-40\% of the healthy cells, and the response dynamics is slow.
If we assume a correlation between internal states, it is
reasonable to look for the fluctuations of different cells.
 According
to [24], the correlation of the Calcium fluctuations in different
mammalian cells decreases rather slowly with the distance $d$
between the cells, like $\sim 1/d^b, b\approx 0.2$, which supports
our hypothesis. Some large scale synchronization has been observed
in the fluctuations of Ca concentration in [13].
 The common sensing of 3-4 cell layers, the correlation between the fluctuations of the Ca-es
concentrations,  and the correlation of Circadian clocks of
thousands cells [25] are three  particular cases of the
correlation between the cells' states in the tissue. We claim that
the correlation is a rather general phenomenon.

\subsection{ The multi-terminal model} We formulated our
hypothesis II about correlation between internal states of the
cells. Now we will formulate a mathematical model  for information
processing by the collection of the cells. Different cells sense
the environment and can also communicate with each other, see Fig.
1b.
 We can replace such a multi-terminal system by another one,
 where the cells just sense the environment. In this case, however,  there is a
 correlation between the states of different cells. We considered
 such a model in [26].
Identifying the information transmission capacity of both systems,
we define the degree of correlation between the states of
different cells.

 We can
qualitatively describe the situation as a decoding of information
via two decoders, related by an almost errorless discrete
information channel. Later, we will look at the realistic case of
the Gaussian channel. We consider a two letter alphabet. Let us
assume that cell A gets the encoded signal, z letters $\pm 1$,
where the probability of the error is $1-p_1$, and original
message about the environment signal has $N$ bits. In the same
way, the second cell B receives a signal of $z$ bits under the
noise probability $1-p_1$. If there is no communication between
the cells, the correct decoding is possible when [12]
\begin{eqnarray}
\label{e2} N\ln 2\le z(\ln 2-h(p_1))
\end{eqnarray}
where $h(p)=-p\ln p-(1-p)\ln (1-p)$ defines the information
content of one letter of the noisy signal as $\ln 2-h(p_1)$, while
original message has $N\ln2 $ bit information.

 We model the collection of the cells as a
multi-terminal system according to Fig. 1.b. Thus, in our system,
there is an information transmission in both directions. Let us
assume that there are communications between the two cells with an
error probability $1-p$ and  cell A sends to cell B just its
signal from the environment. A simple consideration gives the
following condition for the errorless decoding:
\begin{equation}
\label{e3} N\ln 2\le z[(\ln 2-h(p_1))+(\ln 2-h(p_2))]
\end{equation}
where $p_2=p_1p+(1-p_1)(1-p)$.  Thus, our system has a better
information processing ability than single cells and can decode
correctly weaker signals with a smaller $z/N$.

When $p \to 1 $,  our system has the same decoding performance as
the model with strongly correlated (identical) states \cite{sa97}
\begin{equation}
\label{e4} N\ln 2\le 2z[\ln 2-h(p_1)]
\end{equation}
 If we remove the correlation between the terminals ($p\to 1/2$),
the correct decoding of information, and, therefore, the decision
making is impossible [26].
 The loss of correlation between the states of the
cells can play a dramatic role, as information processing in the
cell is near the threshold of correct decoding of noisy
information [27].

According to the experimental results of [28,13], cell-cell
communication is weaker between the cancer cells, the same as with
the correlation of the internal states. Therefore, according to
our interpretation of the cells ensemble (as a multi-terminal
system),  for the cancer case, the multi-terminal system is below
the threshold to construct a normal tissue.

Let us consider another model, when the states of two cells are
correlated,  there are $2^{N(2-C)}$ configurations, less than all
the possible ones $2^{2N}$. Thus, $C$ defines the correlation
level, and $C=1$ is the perfect correlation. These cells sense the
environment.

We get for the error threshold of the model:
\begin{equation}
\label{e5} N\ln 2(2-C)\le z[\ln 2-h(p_1)+\ln 2-h(p_1)]
\end{equation}
Let us identify the error threshold of our model with the error
threshold given by eq. 3. Then we obtain the  effective value of
the correlation $C$.

\begin{figure}
\large \unitlength=0.1in
\begin{picture}(42,12)
\put(-0.5,0.5){\includegraphics{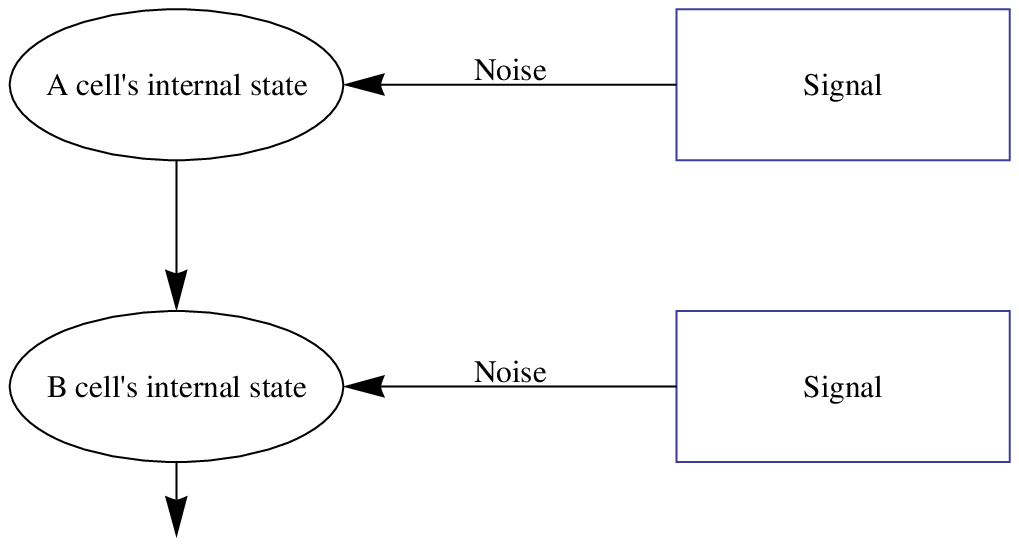}}
\put(16.5,0.5){\includegraphics{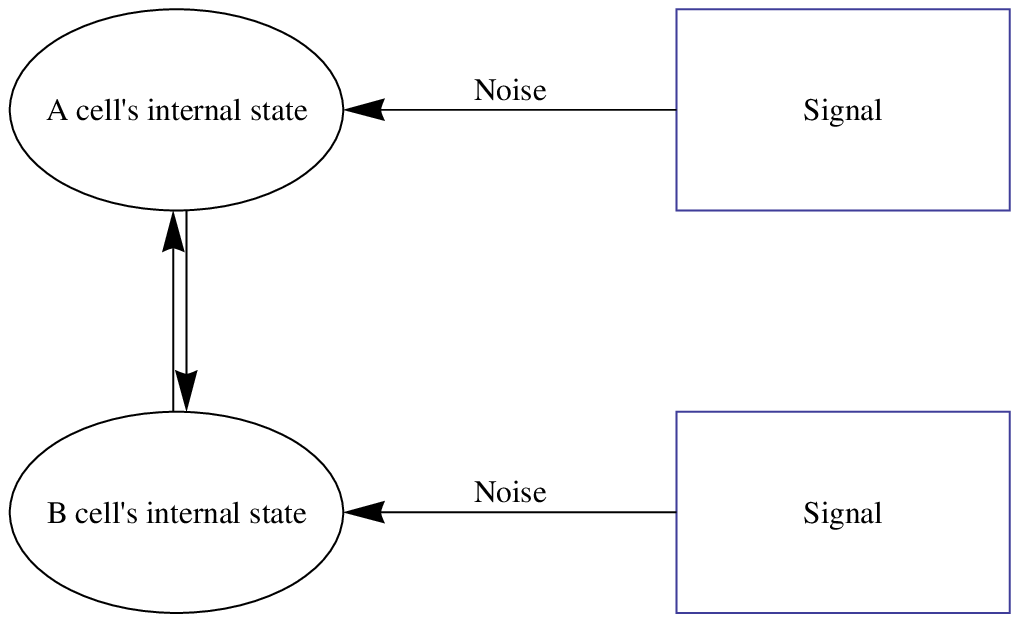}}
\put(7.7,0){\small{a.}} \put(26,0){\small{b.}}
\end{picture}
\caption{Looking the collection of the cells as a multi-terminal
system. The system is decoding the noisy information to make a
decision.  (a). The relay channel from [11].  Both cells sense the
signal from the environment. Cell A sends a signal to the cell B,
which also sense the environment, and gives as output the decoded
information. (b) There is a fast information transition between
two cells, which also senses the environment. }
\end{figure}
A strong correlation means $(1-C)\ll 1$. We see that in order to
get a strong correlation between healthy and cancer cells, we need
 a communication with minimum errors. In the case of errorless
cell-cell communication ($p=1$), there is a complete correlation
between the states of two cells.

We can write the solution of the system of eqs.  3,5 as
\begin{equation}
\label{e6} C= \frac{I_{cc}}{I_{t}}
\end{equation}
where $I_{cc}$ is the information transmission rate between cells
(related to the environment state), and $I_t$ is the total
information transmission rate between the cell, other cells and
the environment (related to the environment state). In case of
Gaussian channel, eq. 6 is equivalent to the constraint that the
signal noise ratio for the communication between neighboring cells
 is much stronger than for the sensing of the
environment.

 While looking at
the correlation between the states of different cells, we should
also  look at
 images of different environmental factors.
 When we put
the cancer cells in a new environment [13], they act as
 normal cells. We can understand the situation in our scheme,
using eq. 6. The information exchange between the cancer cells and
healthy cells is the main fraction in total information
transmission of the cancer cell; therefore, the internal state of
the cancer cell  is still correlated with the states of healthy
cells.

\subsection{ The synchronization and correlation of the states of
cancer cells} We claim that for collective decision making, the
ensemble of the cells should have correlated internal states. For
the cancer cells, the correlation is not enough to organize a
healthy tissue. Nevertheless, it can be enough for some
correlation for a large enough ensemble of the cells. One
possibility  for communication among the cancer cells is through
the diffusive molecules by an all-to-all scheme. Let us regard the
cells  as just oscillators with randomly distributed frequencies.
 Thus, we are characterizing the
internal state of the cell just via a phase. There is a simple
interaction between the oscillators, decreasing with distance.
Consider the Kuramoto model where $N$ oscillators are defined in
the space [29]
\begin{eqnarray}
\label{e7}
\frac{d\phi}{dt}=\omega_i+\frac{K}{\eta}\sum_j\frac{1}{r_{ij}^{\alpha}}\sin(\phi_j-\phi_i),
\end{eqnarray}
where $\omega_i$ are the internal frequencies of the cells and
have some random distribution, $\eta$ is a normalization constant,
and $r_{ij}$ is the distance between the oscillators. For
$\alpha\le 1$, a complete synchronization has been found for the
sufficient large population of the cells. The threshold value of
the population is defined by $\alpha$ and $ K$. Thus, there is a
minimal population size for synchronization. We can qualitatively
understand the growth of the tumor, assuming a minimal size
necessary for synchronization, and therefore, for  collective
decision making by a tumor.
 We claim that a large distance correlation
between the cells' internal states is possible (see also [13]).
Even if the cancer cells communicate weakly with each other, they
can sense the environment signals, and, according to eq. 6, they
can get strong correlation between internal images.

Let us derive a similar result using a multi-terminal model
approach. Different cells sense the environment and communicate
with each other via diffusive molecules, typical for the case of
public goods [30] using the Gaussian information channel [11,12].
Now the error threshold is given by
\begin{equation}
\label{e8} N\ln 2\le z \sum_j\frac{(J_i)^2}{2j^2},
\end{equation}
where $J_i$ is the strength of the signal from the i-th cell to
the given one, while $j$ is the noise intensity. We assume that
due to diffusion $J_i\sim 1/r_i$, where $r_i$ is the distance to
the i-th cell, and there is a uniform distribution of the cells in
the space. The right hand side of eq. 8 grows linearly with the
maximal distance L, and we get from eq. 8
\begin{equation}
\label{e9} (1-C)\sim 1/L,
\end{equation}
where $L$ is the tumor size (diameter).  Thus, the cells' states
are strongly correlated, as in the case of Kuramoto model.


Our result, eq. 9, supports strong cooperation of the cancer
cells. Therefore, while independent replicator models of cancer
are useful at the first stage of disease [31], we have to take
into account the cell-cell interaction later.
 According to [7], a cooperation between the regularity
networks of different cells in tissue can drastically increase the
associative learning ability of cancer cells, $\sim N$ instead of
$\sim \sqrt{N}$ [7] where N is the number of bi-stabilities. The
associative learning should be considered as the main arsenal of
adaptability of cancer cells: short evolution periods in the case
of cancer does not allow more involved schemes of adaptability. We
interpret the results of [6] that the cell-cell cooperation can
increase the adaptability potential by about 10 times. We assume
that this drastic increase in adaptability is possible only
because of the strong correlation of the cells' states.

In [13], it has been assumed that the cell ensemble works near the
critical point (via cell-cell communication parameters) which
allows a control of the cell dynamics via a cell-cell
communication level. We assume that for a proper cooperation of
cancer cells, it is necessary and sufficient to have strong enough
cell-cell communication and metabolism efficiency (to support the
former). Then after further degradation of these features, they
will fail to cooperate. In principle, the cancer cells can
organize a correlation between their internal states using both
diffusive molecules and gap junctions [32], as the latter have
about 60-70 \% of the communication rate of a healthy cell. It is
very important to identify which communication scheme plays a key
role, especially in case of metastasis. Perhaps in this way, we
can understand the metastasis as a new phase where a large scale
cooperation of cancer cells stop to work.

\section{Conclusion}

In conclusion, we derived rigorous mathematical results (four as
an effective dimension of healthy cell division
 dynamics;
a strong correlation of internal states of the cancer cells in the
tumor) as well as suggested several hypothesis to understand the
cell decision making  and cancer which can be verified
experimentally. If we look to the cancer as a complex dynamical
(statistical physics) phenomenon, then the first step should be
the identification of the effective dimension of the phenomenon;
otherwise, a serious investigation of the system is impossible. We
are trying to give a maximally compressed, but still reasonable
understanding of the considered complex phenomena and
 suggest  the comparative analysis of
 healthy and cancer cells by looking at the cell effective dimension
 during the division, the existence of energetic cell-cycle
 related rhythms, the correlation between internal states of the
 cells in tissue, the fractions of gap junction, and
 diffusive molecules in the cell-cell communications.
  We assumed a hypothesis that the rhythms of the cell (circadian and cell-cycle) strongly
 affect the decision making process.
Instead of relating the origin of the cancer with the change of
one rhythm (i.e. a disruption of circadian rhythm; see a critical
review in [17]), we suggest to look at the rhythm-rhythm
interaction.
   We also considered a simple model and derived
 the expression for the correlation of the internal images of
 different cells.
 Our claim is that an efficient collective
 decision making by the cells is possible only by having a strong
 correlation of internal images.
The cancer cells have less effective information processing
ability and that's why they can not support a normal homoeostasis.
 Nevertheless, there is some
 correlation between the cancer cells' states. Assuming just a gaussian property for the cell-cell information channel
 and public goods as the main mechanism of interaction of cancer cells, we derived our
 key result eq. 9 : a strong correlation of the states of the cancer cells.
 Actually, strong correlation of internal states can bring
 self-identification of a group of the cells. We used a simple
 model.
 We need also to look at how the cell-cell
communication errors affect a strong increase of dynamic memory
for associative learning by the ensemble of the cells as found in
[6]. We need to clarify which version of cell-cell communication
(via diffusive molecules or gap junction) is more important in the
case of cancer and metastasis. The metastasis phase can be related
with the transition from long-ranged correlation among the cancer
cells to short range correlation.

 As the rhythms and correlation of internal states of
the cells are crucial in our approach to cancer,
 we suggest that the experiments [10,16] are repeated
 for the cancer case, followed by ATP rhythm looking for
the metastasis, and for the tumor after metabolic treatment [33],
looking for the rhythms and their correlations. Especially
interesting would be  to repeat [13] for the metastasis case,
measuring the cell- cell correlation for calcium in case of the
cancer-metastasis cell and metastasis-metastasis cases. Before
applying the treatment strategy against the public goods, related
to the diffusive molecules [30], we should clarify which version
of communication plays a key role in case of cancer and
metastasis. We should investigate the causal relation between
different rhythms in other living systems \cite{bu09}, as well as
construct a simple models for the system with several rhythms and
decision making. We assume that the cancer's treatment should be
looked at from the both  physical aspects (correcting the
metabolism [33], the acidity [34]) and
 decision making aspects (acting on the rhythms interaction).

\acknowledgments The work was supported by the grant  MOST
104-2811-M-001-102. DBS thanks  R. Amritkar, A. Bratus, F.
Fontanari, S. Jain, H. Qian, J. Pepper,
 A. Martirosyan, J. Prost, B. Sun for the useful discussions and especially
Laurent Schwartz for informing the results of his experiments
before publication in the press.


\begin{thebibliography}{0}

\bibitem{pe77} PETO R.
In: HIATT H. H., Watson J. D., WINSTEN J. A., editors. Origins of
Human Cancer. New York: Cold Spring Harbor Publications;  p.
1403(1977).

\bibitem{no76} NOWELL P. C.  {\it Science}, {\bf 194}, (1976),
23.


\bibitem{me15}AKTIPIS C. A. et al,
Cancer across the tree of life: cooperation and cheating in
multicellularity. Phil. Trans. R. Soc. B {\bf 370}: 20140219
(2015).
\bibitem{ta08} TAGKPOULOS I.,  LIU Y. C.
{\it Science}, {\bf 320}, (2008),1313.

\bibitem{fr12}
FREDDOLINO P. L., TAVAZOIE S.
{\it Annu. Rev. Cell Dev. Biol.}, {\bf 28},(2012) 2.1.

\bibitem{so13}
SOREK M., BALABAN N. Q.,  LOEWENSTEIN Y.
 {\it  PLOS Computational
Biology}, {\bf  9},(2013), e1003179.


\bibitem{we14} WESTERHOFF H. V. et al
{\it Frontiers in Microbiology}  {\bf 5}, (2014)  379.

\bibitem{wo08} WOLF D. M.
 {\it Plos One} {\bf 3}, (2008), e1700.

\bibitem{sa12}SAAKIAN D. B. and SCHWARTZ L.
{\it Euro Phys. Lett.}, {\bf 100}, (2012),68003.


\bibitem{sa15} SANDLER O. et al (2015)
 {\it  Nature}, {\bf 519}, 468.


\bibitem{el15} ELLISON D. et  al
arXiv:1508.04692(2015).
\bibitem{cov} COVER T. M. and  THOMAS J. A., Elements of Information
Theory (Wiley, New York, 1991).
\bibitem{po15}POTTER G. D., BYRD T. A., MUGLER A. and SUN B.
Arxiv 1508.06966(2015).

\bibitem{ba11}BALAZSI G., van OUNDERAARDEN A., and COLLINS J. J. Cellular Decision
Making and Biological Noise: From Microbes to Mammals Gabor, {\it
Cell} {\bf 144}, (2011), 11.

\bibitem{ka09}KADANOFF L.  Reference frame, Phys. Today, {\bf 62}:, (2009), 8.

\bibitem{bi14}
BIELER J. et al,Mol Syst Biol.  {\bf 10}, (2014), 739.
\bibitem{sa14} SANCAR A. et al,
 {\it Biochemistry} {\bf
54}, (2014), 110.




\bibitem{mi11}  MIAN I. S. , ROSE C.
{\it Integrative biology} {\bf 3},(2011) 350.



\bibitem{ge04} GELLe-MANN  M. and  LLOYD S.
edited by M. Gell-Mann and C. Tsallis,  Oxford University Press,
New York, (2004) p. 387.

\bibitem{sa05} SAAKIAN D. B.
  {\it Phys. Rev. E}, {\bf 71},(2005), 016126.
  \bibitem{ku98} SCHAFER C., ROSENBLUM M. G., KURTHS J., ABEL H. H.
   {\it  Nature} {\bf 392}, (1998) 239.

\bibitem{bu09} BUCHNER T. et
al (2009)
 da S.K DANA et al.
(eds.), Complex Dynamics in Physiological Systems: From Heart to
Brain, Understanding Complex Systems, p.33, Springer.
\bibitem{ta10}  TAY  S. al
 {\it Nature}, {\bf 466},(2010)267.

\bibitem{su13}SUN B., DUCLOS G., and STONE H. A.,
Phys. Rev. Lett. {\bf 110}, (2013), 158103.
\bibitem{ul09} ULLNER E., et al
Biophysical Journal {\bf 96},(2009), 3573.


\bibitem{sa97} ALLAKHVERDYAN A. E., SAAKIAN D. B.
JETP {\bf 111}, (1997), n.3.



\bibitem{bo14}BOWSHER C. G. and  SWAIN P. S.
 {\it
Current Opinion in Biotechnology} {\bf 28}, (2014), 149.

\bibitem{lo66}LOEWENSTEIN W. R.,  KANNO Y.,
{\it  Nature} {\bf 209}, (1966)1248.




\bibitem{ma02} MARODI M.,
d'OVIDIO F., and VICSEK T.
{\it Phys. Rev E} {\bf 66},(2002), 011109.
\bibitem{pe14}PEPPER J.,
{\it  Evolution, Medicine, and Public Health} pp. 6568
doi:10.1093/emph/eou010(2014) .
\bibitem{no10} BOZICA I. et al
 {\it PNAS} {\bf  107}, (2010), 18545.
\bibitem{su12} SUN B. et al
{\bf 109}, (2012)7753.

\bibitem{la13} ABOLHASSANI M.
 et al
Invest New Drugs, DOI 10.1007/s10637-011-9692-72013 (2013).

\bibitem{lev}  CHERNET B. T. and LEVIN M.
{\it  Dis. Model. Mech.} {\bf 6},(2013),595.



\end{thebibliography}
\end{document}